\documentclass[longbibliography,floatfix,reprint,twocolumn,notitlepage,prb,superscriptaddress,showpacs]{revtex4-2}

\usepackage{hyperref}
\usepackage[usenames,dvipsnames]{color}
\usepackage{amsmath}
\usepackage[english]{babel}
\usepackage{amssymb}
\usepackage{graphicx}
\usepackage{epstopdf}
\usepackage{comment}
\usepackage{soul}
\usepackage{marvosym}
\usepackage{mathrsfs}
\usepackage{amsfonts}
\usepackage{tabularx}
\usepackage{multirow}
\usepackage{xcolor}
\usepackage{amsmath}
\usepackage{soul}
\usepackage{makecell}
\usepackage{mathtools}

\usepackage[normalem]{ulem}
\usepackage{wasysym}

\let\oldbibitem\bibitem
\renewcommand{\bibitem}{%
  \renewcommand{\doi}[1]{doi: ##1}
  \let\bibitem\oldbibitem
  \oldbibitem
}

\begin{document}

\preprint{APS/123-QED}

\title{Concentration Dependence of Elastic and Viscoelastic Properties of Aqueous Solutions of Ficoll and Bovine Serum Albumin by Brillouin Light Scattering Spectroscopy}

\author{Stephen J. Spencer, Venketesh Thrithamara Ranganathan, Anand Yethiraj, and G. Todd Andrews}
 
\affiliation{
 Department of Physics and Physical Oceanography, Memorial University, St. John's, NL, A1B 3X7, Canada
}

\date{\today}

\begin{abstract}

The cellular environment is crowded with macromolecules of different shapes and sizes. The effect of this macromolecular crowding has been studied in a variety of synthetic crowding environments: two popular examples are the compact colloid-like Ficoll macromolecule, and the globular protein bovine serum albumin (BSA). Recent studies have indicated a significant component of bound or surface-associated water in these crowders reduces the available free volume. In this work, Brillouin light scattering experiments were performed on aqueous solutions of Ficoll 70 and Ficoll 400 with concentrations ranging from 1 wt\% to 35 wt\% and BSA with concentrations of 1 wt\% to 27 wt\%. From the dependence of spectral peak parameters on polymer concentration, we determined fundamental solution properties: hypersound velocity, adiabatic bulk modulus and compressibility, apparent viscosity, and hypersound attenuation. Existing theory that ignores intermolecular interactions can only capture the observed linear trends in the frequency shift up to a threshold concentration, beyond which a quadratic term accounting for intermolecular interactions is necessary. This likely indicates a transition from the dilute to semi-dilute regime. In the Ficoll solutions (but not BSA) we see evidence for a central mode, with a characteristic relaxation time of 20 ps, that we attribute to exchange of the bound water.
\end{abstract}

\pacs{Valid PACS appear here}
\maketitle

\section{Introduction}
Physical systems consisting of liquid water and macromolecules are ubiquitous. The fluid medium inside a biological cell is an aqueous solution that consists of macromolecules of different sizes and shapes which occupy a significant volume (typically assumed to be 30 - 40$\%$ of the cell) \cite{ellis2001macromolecular}. It has been understood for more than two decades that the crowded macromolecular environment can affect biochemical reactions within the cell~\cite{Minton2001}. Any volume other than the water volume is inaccessible to other molecules, and this excluded volume affects molecular structure, motions and chemical kinetics. In addition, for compact molecules with internal bound water, the accessible water volume will be less than the total water volume. Thus, the simplest picture of macromolecular crowding is entropic: the macromolecules and the inaccessible water reduce the free volume and increase the excluded volume.

While there is an increasing realization of the role of other non-specific interactions~\cite{Rivas2022}, experimental model systems have focused on aqueous solutions with a simple crowder such as a polysaccharide ({\it{e.g.}}, the compact Ficoll macromolecule or the chain-like dextran) or a globular protein
({\it{e.g.}}, bovine serum albumin) \cite{dhar2010structure, acosta2017large, norris2011true}. Structure and dynamics in these crowders has been reported extensively \cite{venturoli2005ficoll, fissell2007ficoll, fodeke2010quantitative, christiansen2013quantification, shah1995adsorptive}. More realistic, heterogeneous crowding media, such as bacterial cell lysate, have been employed as well \cite{sarkar2014protein,trosel2023diffusion}. In recent work, Ranganathan $et$ $al$. \cite{rang2022} have provided quantitative evidence, via pulsed field gradient nuclear magnetic resonance spectroscopy, for water that is bound and thus inaccessible to other molecules, implying that the true excluded volume is larger than what is usually inferred. 
 
Brillouin light scattering spectroscopy has been recognized as a niche technique for probing the mechanical and viscous properties in heterogeneous biomaterials \cite{Palombo2019}. Brillouin scattering experiments on aqueous macromolecular solutions and hydrogels typically report the dependence of spectral peak parameters and derived elastic and viscoelastic properties on solute concentration \cite{adic2019,tao1989reorientational,tao1993light}. These parameters and properties are usually measured over a large concentration range and include Brillouin peak frequency shift and linewith, hypersound velocity and attenuation, apparent viscosity, and various elastic and viscoelastic moduli.  Attempts at using existing theoretical models to describe the concentration dependence of these quantities, however, have met with limited success.  For example, Brillouin studies of so-called ``aqueous biorelevant solutions'' \cite{adic2019} reveal a common dependence of peak frequency shift on solute concentration for those solutions containing macromolecules (lysozyme, bovine serum albumin, and gelatin) up to 40 wt\%, implying that this behaviour is largely independent of the nature of the solute.  Application of the Reuss effective medium model for a two-component system resulted in an equation for the shift as a function of concentration that agreed well with most experimental data. A theoretical expression relating peak linewidth to concentration that incorporated this equation, however, was unable to fully reproduce the observed trends.  Moreover, the sparsity and relatively large concentration interval ($\sim10$ wt\%) between adjacent data points means that possible discontinuities in the observed trends, such as might occur in proximity to the polymer overlap concentration, were not accounted for in the model.  In related Brillouin scattering work on collagen hydrogels \cite{bail2020} the storage modulus of collagen was extracted by fitting an expression for the concentration dependence of the effective storage modulus from the Voigt model to a high-hydration subset of the full experimental dataset for the concentration dependence of the gel storage modulus.  It was stated that the two-component Voigt model gave a better fit than the Reuss model used in previous works \cite{bail2020}.  No model was advanced to describe the data trend over the entire measured concentration range.  Moreover, Brillouin scattering studies of cross-linked polyvinyl alcohol hydrogels \cite{ng1985} found only crude qualitative agreement between the observed dependence of Brillouin peak frequency shift and linewidth on gel network volume fraction and that calculated from a theory incorporating frictional damping and coupling between elastic waves in polymer network and fluid \cite{marqu1981}.

In this paper we report on Brillouin light scattering studies of aqueous solutions of polymers Ficoll 70, Ficoll 400, and Bovine Serum Albumin (BSA).  Solution elastic and viscoelastic properties were determined over the dilute and semi-dilute ranges from the dependence of spectral peak parameters on solute (polymer) concentration. The observed trends in these properties are not consistent with existing theory but instead were found to be well-described by expressions derived from a new model relating hypersound frequency and solute concentration.  The sensitivity of the Brillouin scattering technique to changes in structure and water-macromolecule dynamics also allowed the polymer overlap concentration and the relaxation time associated with the hydration of Ficoll molecules to be determined. The extent of the hydration, manifested in the unexpectedly low overlap concentration, provides independent confirmation of recent results that suggest that the effective volume fraction occupied by hydrated macromolecules in solution is much larger than expected for the bare unhydrated variety. In characterizing the viscoelastic properties of commonly-used experimental model systems over a wide solute concentration range and advancing a model that incorporates interparticle interaction, this study provides important new insight into the physics of macromolecular crowding and biomacromolecular systems in general.

\section{Experimental Details}
\subsection{Sample Preparation}
Solutions of Ficoll-70 ($m=70$ kDa) and Ficoll-400 ($m = 400$ kDa) were prepared under identical conditions by dissolving Ficoll powder in 99.9\% pure D$_{2}$O at room temperature in small glass vials. A Scientific Industries Vortex Genie was used to promote initial mixing of D$_{2}$O and Ficoll, and subsequent homogenization was performed using a Fisherbrand Homogenizer 850 with a rod diameter of 7 mm and speed of 11,000 RPM. The homogenization sequence, which was performed five times per sample, consisted of three minutes of mixing followed by one minute of settling.  The resulting solutions were clear and colourless and had concentrations ranging from 1 wt\% to 35 wt\%, with a noticeable increase in viscosity for those at the upper end of this range.

BSA solutions with concentrations of 1\% w/w to 27\% w/w were prepared by dissolving BSA powder in 0.1 M phosphate buffer solution with pH of 7.0 at room temperature and subjecting them to the same mixing and homogenization procedure as used for the Ficoll solutions.  Solutions with concentrations $< 20$ \% w/w were clear and colourless while those with concentrations higher than this value were somewhat cloudy. 

\subsection{Brillouin Light Scattering Spectroscopy}
\subsubsection{Apparatus}
Brillouin light scattering experiments were performed under ambient conditions using a 180$^{\circ}$ backscattering geometry. The incident light source was a Nd:YVO$_{4}$ solid state laser with an emission wavelength of 532 nm and output power of 1.66 W. To minimize Fresnel reflection losses, a half-wave plate was used to rotate the plane of polarization from vertical ($s$-polarized) to horizontal ($p$-polarized).  Neutral density filters placed in the beam path were used to reduce the power level at the sample to $\sim100$ mW.  Light was focused on samples with a 5 cm lens of $f$-number 2.8. Scattered light was collected and collimated by the same lens and subsequently focused by a 40 cm focal length lens onto the 450 $\mu$m entrance pinhole of a six-pass tandem Fabry-Perot interferometer for spectral analysis.  The interferometer had a free spectral range of 15 GHz and a finesse of $\sim100$.  A schematic of this experimental setup can be found in Refs. \cite{andrews2007brillouin, andr2018}. 

\subsubsection{Quantities Derived from Spectra}
Elastic and viscoelastic properties of the Ficoll solutions were deduced from Brillouin spectra.  Hypersound velocity was determined using the well-known Brillouin equation applied to the case of a backscattering geometry, 
\begin{equation}
v=\frac{f\lambda}{2n},
\label{eq:fvel}
\end{equation}
where $f$ is the measured Brillouin peak frequency shift, $\lambda$ is the incident light wavelength, $n = n(x)$ is the concentration-dependent solution refractive index, and $x$ is the concentration in weight percent.  The latter was obtained for each Ficoll 70 and Ficoll 400 solution using $\partial n/\partial x$ relationships provided in the literature \cite{fissell2010size}, however this was found to be constant at $n=1.33$ for the full concentration range. This, however, was not the case for solutions of BSA. At higher concentrations of BSA, solutions became noticeably cloudy. As such, refractive indices for BSA solutions were calculated using the relationship $\partial n / \partial C = 0.190$ mL/g (with C expressed as g/mL), found by  Tumolo $et$ $al$. \cite{tumolo2004determination}.

Knowledge of the hypersound velocity allowed the adiabatic bulk modulus 
to be found from
\begin{equation}
    B=\rho v^2,
    \label{eq:fbulk}
\end{equation}
where $\rho$ is the mass density of the solution. The density of the solution showed little variation over the full concentration range, and was approximated to a constant 1110 kg/m$^{3}$ for the purposes of fitting. The adiabatic compressibility, $\kappa$, was also  determined using the fact that $\kappa = 1/B$. 

The apparent viscosity, $\eta$, and hypersound attenuation, $\alpha$, in the solution were deduced from Brillouin spectral data via
\begin{equation}
\label{eq:fvisc}
\eta = \frac{4}{3}\eta_s + \eta_b = \frac{\rho v^{2} \Gamma_B}{4 \pi^{2} f^{2}} = \frac{\rho \lambda^2 \Gamma_B}{16 \pi^2 n^2}
\end{equation}
and
\begin{equation}
     \alpha = \frac{\pi \Gamma_B}{v},
     \label{eq:fabs}
\end{equation} 
respectively \cite{ostwald1977high}, where $\Gamma_B$ is the Brillouin peak full width at half-maximum (FWHM), $\eta_s$ and $\eta_b$ are the shear and bulk viscosities, and the other quantities are as already defined.

\section{Results and Discussion}

\subsection{Spectra}

\subsubsection{General Features and Mode Assignment}
\begin{figure}
    \centering
    \includegraphics[scale=0.3]{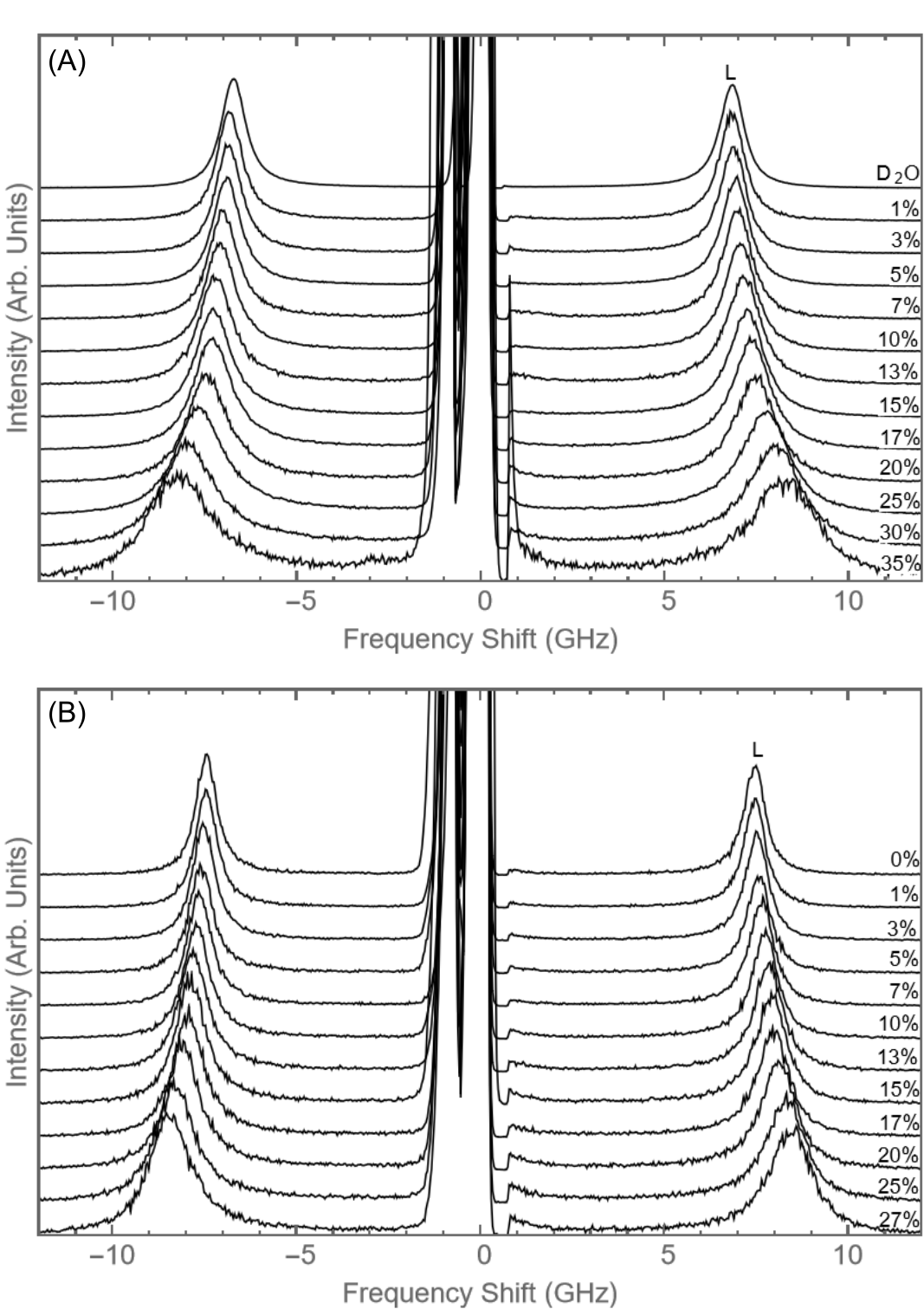}
    \caption{Normalized Brillouin spectra collected from solutions of (A) Ficoll 70 and (B) BSA of various concentrations (wt\%). L represents a longitudinal bulk mode. }
    \label{fig:f70stacked}
\end{figure}

\begin{figure}
    \centering
    \includegraphics[scale=0.3]{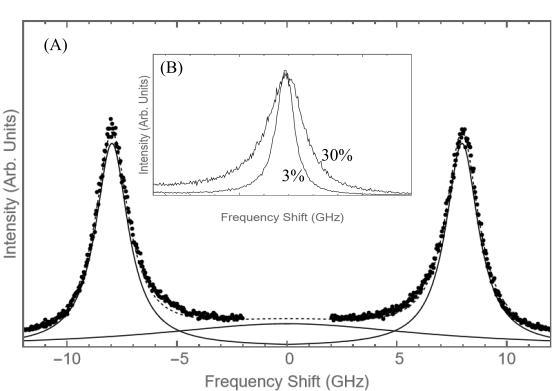}
    \caption{(A) Brillouin spectrum of an aqueous solution of Ficoll 70 with a solute concentration of 30\%. Solid lines - Best-fit Lorentzian functions for central peak and Brillouin peaks. Dotted line  - sum of central peak and Brillouin peak best-fit Lorentzians. (B) Anti-Stokes Brillouin peaks for aqueous Ficoll solutions with solute concentrations of 3\% and 30\%. 30\% concentration peak is shifted horizontally from 7.96 GHz to 6.86 GHz so that the peaks overlap to highlight the slight asymmetry and significantly higher baseline intensity of the peak for the 30\% concentration solution compared to that for the 3\% solution on the low frequency shift side.}
    \label{fig:fasymm}
\end{figure}

Figure \ref{fig:f70stacked} shows a series of Brillouin spectra collected from the Ficoll 70 solutions and BSA solutions. Spectra of the Ficoll 400 solutions are similar to Ficoll 70 spectra in all respects (see Figure S1 in the Supplementary Material).  A single set of Brillouin peaks was observed in all spectra, with a frequency shift ranging from $\sim6.8$ GHz to $\sim8.4$ GHz.  Although not obvious from the spectra shown in Figure \ref{fig:f70stacked}, there is also a broad, weak peak at the center of the spectra obtained from Ficoll 70 and Ficoll 400 solutions with solute concentrations $\geq 20$\%.  The presence of this peak was inferred from the fact that the baseline intensity in the region between the central elastic peak and the Brillouin doublet in spectra of high concentration solutions was noticeably higher than that in spectra of the low concentration solutions and also higher than that on the high frequency shift side of the Brillouin doublet.  An example of this for Ficoll 70 solutions with 3\% and 30\% concentration is shown in Figure \ref{fig:fasymm} A. In contrast, no central peak was discernible in spectra of the BSA solutions. 

The Brillouin doublet and the peak in the center of the spectra of high concentration Ficoll solutions have different origins.  The former is assigned to the usual longitudinal acoustic mode propagating through the solution based on the similarity of its frequency shift to that of the corresponding mode in water \cite{xu2003measurement}.  The central peak was attributed to a diffusive relaxation mode based on its zero frequency shift and the fact that the width of the peak showed no significant change with changing concentration \cite{pinnow1968rayleigh}. These properties are consistent with other Brillouin and Rayleigh scattering studies which have observed this relaxation mode in macromolecular solutions. \cite{tao1989reorientational,tao1993light,lee1993brillouin,poch2006struc}

\subsubsection{Extraction of Peak Parameters}
Longitudinal acoustic mode peak frequency shift and linewidth were obtained by fitting Lorentzian functions to the Brillouin peaks, with the latter being processed prior to plotting by subtraction of the 0.3 GHz instrumental contribution to the best-fit linewidth.

To obtain an estimate of the central mode linewidth, $\Gamma_C$, it was first necessary to remove data from regions of the spectrum containing other peaks so as to minimize its impact on any subsequent fit.  This included the region containing the central elastic peak and other data contained within the central Fabry-Perot interferometer control window, $\pm2$ GHz from the center of the spectrum, and that containing the two longitudinal mode peaks, which was typically a $\sim$ 6 GHz range around the longitudinal peak.  The remaining data was fitted to a Lorentzian function (see Fig. \ref{fig:fasymm}A), revealing a central mode that is very weak and exceptionally wide, with a FWHM of $\sim14$ GHz.    As can be seen by the high degree of overlap between the experimental data and the dotted curve in Fig.~\ref{fig:fasymm}A, the addition of this Lorentzian to the best-fit Lorentzians for the longitudinal mode peaks results in a function that well represents the original Brillouin spectrum (without the central elastic peak).

\subsection{Longitudinal Acoustic Mode: Elastic and Viscoelastic Properties}
\subsubsection{Dependence of Brillouin Peak Parameters on Solute Concentration}
\begin{figure}
    \centering
    \includegraphics[scale=0.28]{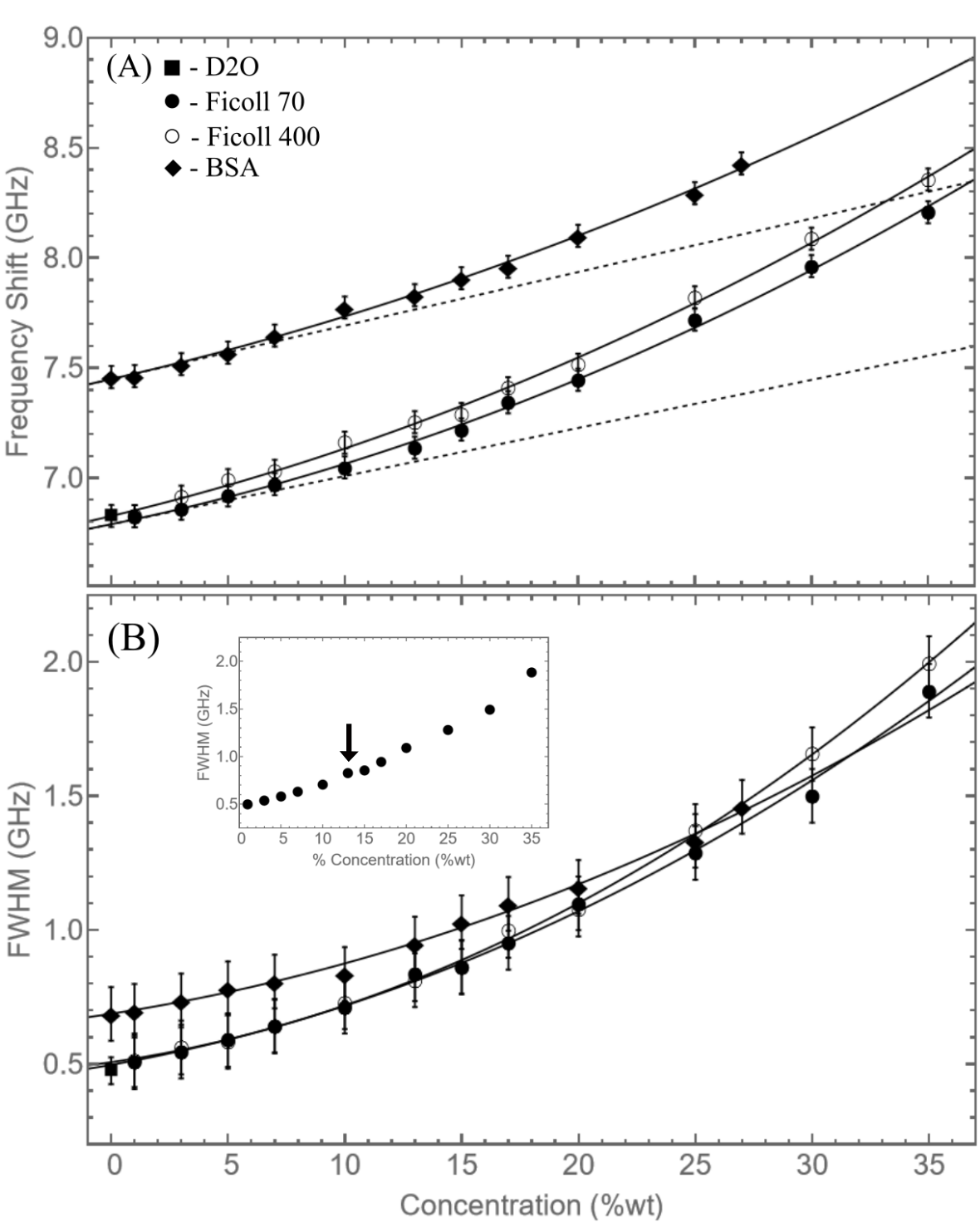}
    \caption{Brillouin peak frequency shift (A) and linewidth (B) as a function of concentration for aqueous solutions of Ficoll 70, Ficoll 400, and BSA. Solid lines represent best fits of $f(x) = f_R(x) + A_1x^2$ and $\Gamma_B(x) = Af(x)^2 + Bx + A_2x^2$. Dashed lines represent frequency relationship provided by Equation \ref{eq:adic1} for Ficoll 70 and BSA. Inset in (B) is linewidth of Ficoll 70.}
    \label{fig:ffreqFic}
\end{figure}

Figure \ref{fig:ffreqFic} shows longitudinal acoustic mode peak frequency shift and linewidth versus concentration for the Ficoll 70, Ficoll 400, and BSA solutions. For all solutions, both quantities increase monotonically with increasing concentration, with the linewidth being much more sensitive than the shift to changes in concentration.  While the frequency shift for each solution increases by $\sim20$\% over the range probed, the peak FWHM for the Ficoll and BSA solutions increases by a factor of $\sim4$ and $\sim2$, respectively. 

For a given concentration, the frequency shift for Ficoll 400 solution is slightly larger than that for Ficoll 70 solution.  At low concentrations, the peak linewidths obtained from the Ficoll 70 and Ficoll 400 solutions are nearly equal, while for concentrations in excess of $\sim15$ wt\% the linewidths begin to diverge, with that for Ficoll 400 being greater than that for Ficoll 70 over this range. This subtle change in the relationship between concentration and frequency or linewidth may represent the overlap concentration at which the system physics changes as it transitions from the dilute to the semi-dilute regime.

The solute concentration dependence of Brillouin peak frequency shift of some so-called biorelevant aqueous solutions have been well-fit for concentrations up to $x \sim40$ wt\% by the relation \cite{adic2019} 
\begin{equation}
    f_R(x) = \frac{f_w}{\sqrt{1-x+xv^{2}_{w}/v^{2}_{s}}} = \frac{f_w}{\sqrt{1+\alpha x}},
    \label{eq:adic1}
\end{equation}
where $\alpha \coloneqq v^{2}_{w}/v^{2}_{s}-1$. Also, the concentration dependence of linewidth has been well fit up to $x\sim20$ wt\% by
\begin{equation}
    \Gamma^B_R(x) = Af_R(x)^2 + Bx , \label{eq:adic2}
\end{equation}
respectively, where $A$ and $B$ are constants and $f_w$, $v_w$, and $v_s$ are the Brillouin peak frequency shift for water, and the hypersound velocities of water and the solute, respectively.  The basis of  Equation \ref{eq:adic1} is the two-component Reuss model for which the effective elastic modulus $M$ of the solution is given by

\begin{equation}
\frac{1}{M} = \frac{\rho \mu_s}{\rho_s}\left[\frac{1}{M_s}-\frac{1}{M_w}\right] + \frac{1}{M_w}, \label{eq:reuss1}
\end{equation}
where $\mu_s = m_s/(m_s+m_w)$ is the solute mass fraction and $M_s$ and $M_w$ are the elastic moduli of the solute and water, respectively.  If the density of the solution $\rho$ equals that of the solute $\rho_s$, then Equation \ref{eq:reuss1} simplifies to
\begin{equation}
\frac{1}{M} = \frac{\mu_s}{M_s}+\frac{1- \mu_s}{M_w}, \label{eq:reuss2}
\end{equation}
Equation \ref{eq:adic1} can be obtained by from Equations \ref{eq:fvel} and \ref{eq:reuss2} with the approximation of fixed density and refractive index.

Fitting Equation \ref{eq:adic1} to the $\{f,x\}$ data for the Ficoll and BSA solutions with $v_{w}^2/v_{s}^2$ as an adjustable parameter, however, yielded best-fit relations $f(x)$ that show only mediocre agreement with experiment - overestimating $f$ at low concentrations and underestimating it at higher concentrations over the range probed.  Substitution of these ``best-fit" expressions for $f(x)$ into Equation \ref{eq:adic2} give similar quality fits to the $\{\Gamma_B,x\}$ data shown in Figure \ref{fig:ffreqFic}.  In addition, a fit of Equation \ref{eq:adic1} to only the low concentration data yields an excellent fit for $x\leq10$\% but the resulting function does not describe the higher concentration data (see dotted curve in Figure \ref{fig:ffreqFic}).  This sub-optimal agreement between theory and experiment is likely due to it not properly accounting for molecular crowding arising from hydration of Ficoll or BSA as the solution concentration increases, the onset of which occurs at the transition from the dilute to semi-dilute regime.  This crowding leads to increased polymer-polymer interaction in the solution.  As such, the hypersound frequency can no longer be described by Equation \ref{eq:adic1}; there are also contributions from volumetric and entropic changes due to polymer-polymer interactions, specifically the onset of contact and the possible formation of loosely packed regions of solute molecules.

The failure of the above model to accurately reproduce the concentration dependence of $f$ and $\Gamma_B$ for the Ficoll and BSA solutions, coupled with the lack of other appropriate theoretical models, lead us to propose a new model to describe the current experimental data.  This model assumes that the concentration dependence of $f$ and $\Gamma_B$ both increase smoothly with concentration according to
\begin{equation}
    f(x)= f_R(x) + A_1x^2,
\label{eq:freqshift}
\end{equation}

where $f_{R}$ is given by Equation \ref{eq:adic1}, and
\begin{equation}
    \Gamma_B(x) = Af(x)^2 + Bx + A_2x^2, \label{eq:FWHM}
\end{equation}
where the $A_i$ are fit parameters and $f(x)$ is given by Equation \ref{eq:freqshift}. Furthermore, the second order term in Equation \ref{eq:FWHM}, $A_{2}x^{2}$ is a phenomenological extension to van’t Hoff’s law, as explained below \cite{rodenburg2017van,al2015virial}. From this, solute molecules in a dilute solution may be treated as an ideal gas. As such, the linewidth is proportional to solution density, as expressed in the equation
\begin{equation}
    \Gamma_B(x) \propto 1 + c_{1} \rho + c_{2} \rho^{2} + c_{3} \rho^{3} + \cdots.
    \label{eq:vant}
\end{equation}
At low concentrations, corresponding to lower density, the higher order terms are insignificant. At higher concentrations, however, these higher order terms begin to dominate \cite{al2015virial}. The second order term of the virial equation is attributed to interactions between solute molecules, the contribution of which becomes more important at higher solute concentrations.

Figure \ref{fig:ffreqFic} shows best-fits of Equations \ref{eq:freqshift} and \ref{eq:FWHM} to the $\{f,x\}$ and $\{\Gamma_B,x\}$ data for the Ficoll and BSA solutions.  The fits reproduce the trends of the experimental data very well.  For reference, the best-fit equations are given in Table \ref{tab:elas_vscoel_eqs}.  

Although not obvious, there is also what could be a subtle change in behaviour which is visible in both the frequency shift and linewidth, which is localized around $\sim15$ wt\% concentration for both Ficoll solutions (denoted by the arrow in the inset in Figure \ref{fig:ffreqFic} for change in behaviour of Ficoll 70 peak width).  The subtle change in behaviour of frequency and linewidth around $\sim$15\% solute concentration is also observed in the BSA data. This change in behaviour oversver may also be indicative of the overlap concentration, and the transition from the dilute to semi-dilute regime. 

\subsubsection{Hypersound Velocity}
\begin{figure*}
    
    \hspace*{-0.5 cm}
    \includegraphics[scale=0.28]{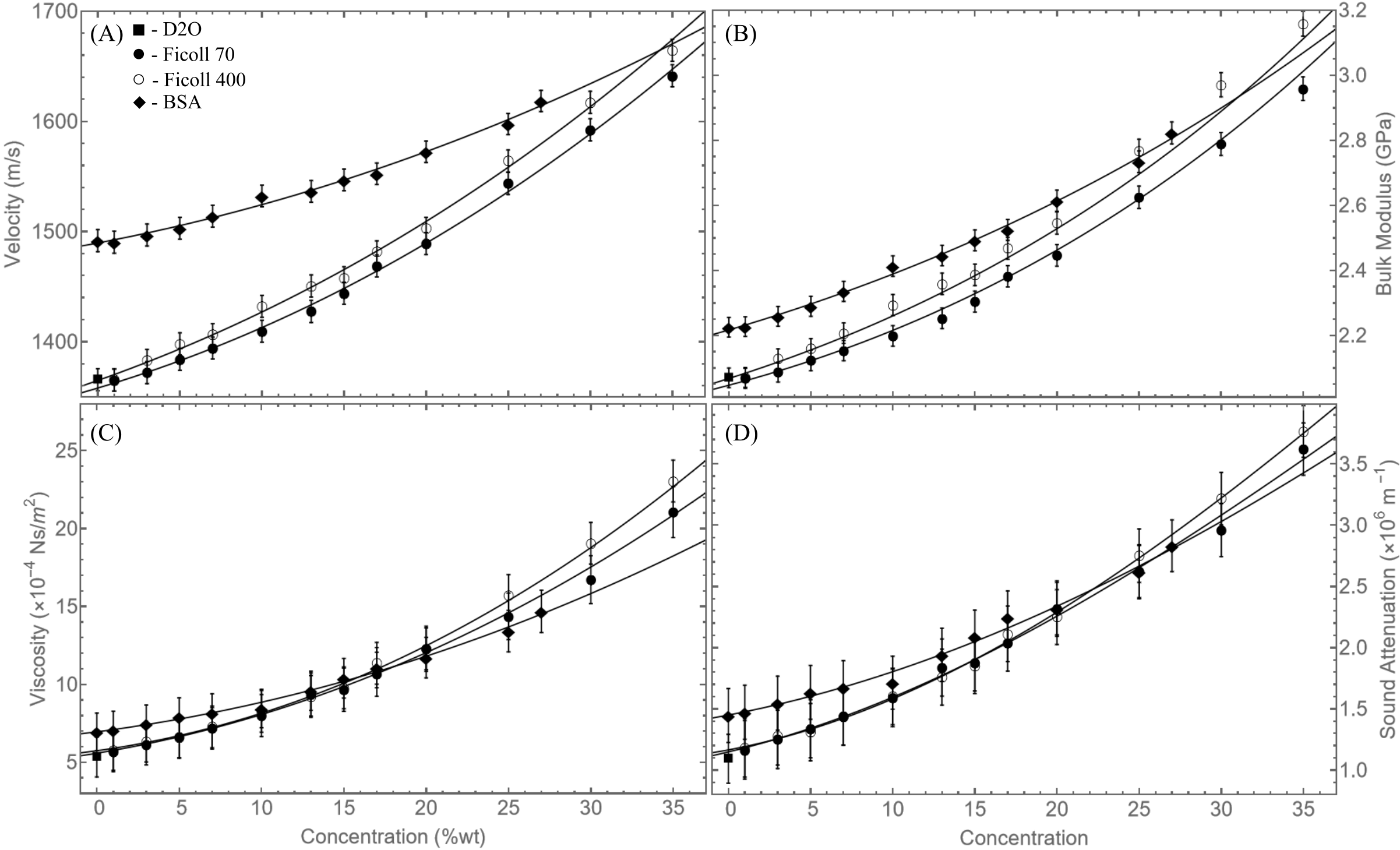}
    \caption{Hypersound velocity (A), solution bulk modulus (B), apparent viscosity (C), and hypersound attenuation (D) as a function of concentration in aqueous solutions of Ficoll 70, Ficoll 400, and BSA. Solid lines are curves are based on the frequency fits from Figure \ref{fig:ffreqFic} and Equations \ref{eq:fvel}, \ref{eq:fbulk}, \ref{eq:fvisc}, and \ref{eq:fabs}, respectively}
    \label{fig:viscoelas}
\end{figure*}

Figure \ref{fig:viscoelas} (A) shows the evolution of hypersound velocity with concentration for the Ficoll 70, Ficoll 400, and BSA solutions. There is a consistent increase in velocity with increasing solute concentration for all solutions, as expected from the increase in Brillouin peak frequency shift with increasing concentration. As with the frequency data, the hypersound velocity in Ficoll 400 solutions is systematically higher than in Ficoll 70 solutions. Furthermore, while the velocity of BSA solutions being larger than that of Ficoll solutions can be attributed to the different solvent, the change in velocity with respect to concentration was much lower for BSA than for both Ficoll solutions. Figure \ref{fig:viscoelas} (A) shows curves for velocity as a function of concentration which were derived by substituting the best-fit equation for frequency as a function of concentration into Equation \ref{eq:fvel}.

In the dilute regime, the hypersound velocity of macromolecular solutions can be expressed in a manner similar to Equation \ref{eq:adic1}, as hypersound velocity is linearly proportional to phonon frequency \cite{adic2019}. However, as the concentration increases beyond 10\%, we begin to see a significant increase in packing of solute molecules in solution due to crowding effects. This is demonstrated by the deviation of Brillouin peak frequency and linewidth from previous theory. It also is important to note that previous studies have shown that solute hydration leads to volume fraction of solute being much larger than expected than expected for bare (no bound water) solute molecules at similar weight concentrations.\cite{rang2022}. This concept is further supported by Brillouin scattering results in this work, as the onset of polymer-polymer interactions was seen at concentrations as low as 10\% by weight. This increase in macromolecular packing leads to an increase in interactions between solute molecules \cite{pethrick1982acoustic}. This is the transition between the dilute and the semi-dilute limits.

By using the $f_{R}(x)$ expression of Equation \ref{eq:freqshift}, the hypersound velocity for Ficoll 70, Ficoll 400, and BSA was calculated. Values for these velocities are shown in Table \ref{tab:solvel}. Velocities calculated for solutes used in this work ranged between $\sim2300$ m/s and $\sim2900$ m/s, which is comparable to results from previous studies on macromolecular solutions \cite{adic2019}.

\begin{table}[t]
\caption{Hypersound velocities for Ficoll 70, Ficoll 400, and Bovine Serum Albumin (BSA) solute, calculated using the equation $v{_w}^2/v{_s}^2-1=\alpha$, where $\alpha$ is the fit parameter in the denominator of $f_{R}(x)$ from equation \ref{eq:adic1}. }
\begin{ruledtabular}
\begin{tabular}{cc}
 Solute & Velocity (m/s)\\ \hline
\multirow{3}{*}{} Ficoll 70 &2320\\
Ficoll 400 & 2850\\
BSA & 2580\\
\end{tabular}
\end{ruledtabular}
\label{tab:solvel}
\end{table}

\subsubsection{Bulk Modulus \& Adiabatic Compressibility}
Figure \ref{fig:viscoelas} (B) shows that the bulk modulus for the Ficoll 70, Ficoll 400, and BSA solutions increase with increasing solute concentration.  This trend was expected not only from the relationship between bulk modulus and hypersound velocity given by Equation \ref{eq:fbulk}, but also from an intuitive perspective as with an increase in concentration a greater proportion of the available volume is occupied by the solute. This leads to the solution being less compressible and therefore to a higher bulk modulus. For low concentrations $B$ increases approximately linearly with $x$ as predicted by Equation \ref{eq:reuss1} in the limit of small $x$. While there is no obvious change in trend, at $x \sim15$ wt\% the slope of $B(x)$ is noticeably larger and continues to increase with increasing concentration.  Following the reasoning in Sec. III.B.1, this is caused by the solution transitioning from dilute to semi-dilute.  The increased interaction between solute molecules ({\it e.g.}, entanglement) results in a decrease in compressibility and, consequently, an increase in bulk modulus.  This deviation of the solution compressibility from a volumetric average of the compressibilities of the constituents ({\it i.e.}, the Reuss model) can be attributed to entropic and volume contributions due to these interactions as discussed in Sec. III.B.1 \cite{pethrick1982acoustic}. 

\subsubsection{Apparent Viscosity}
The apparent viscosity of the Ficoll and BSA solutions for a range of concentrations in the dilute and semi-dilute regimes are shown in Figure \ref{fig:viscoelas} (C).  The $\eta(x)$ curves in this figure were obtained from Equation \ref{eq:fvisc} using $\Gamma_B(x)$ given by Equation \ref{eq:adic2} with the best-fit $f(x)$ from Equation \ref{eq:fvisc}.  As can be seen, the apparent viscosity values for the two Ficoll solutions are roughly equal at low concentrations but begin to diverge from eachother at $x\sim15$ wt\%, behaviour similar to that observed for relative viscosity in rheology studies of aqueous Ficoll 70 and Ficoll 400 solutions \cite{rang2022}.  This change in behaviour approximately at a concentration corresponding to the dilute-to-semi-dilute transition.  Within this region, the contribution to viscosity from Ficoll becomes more dominant than that of D$_{2}$O. It is important to note that previous rheology studies have shown that the intrinsic viscosities of Ficoll 70 and Ficoll 400 differ by $\sim 12\%$ \cite{rang2022}. Furthermore, the range of apparent viscoties presented in this work are significantly different the ranges of shear viscosities shown in previous rheology studies. While the very low concentration viscosities are comparable, the shear viscosities presented in previous rheology work is larger than the apparent viscosities drerived from Brillouin scattering by a factor of $\sim 100$ \cite{rang2022}. At low concentrations, apparent viscosity of BSA solutions is slightly larger than that of either solution of Ficoll. The curve of Equation \ref{eq:fvisc} for BSA, however, is considerably less steep than that of either solution of Ficoll, demonstrating a lower intrinsic viscosity for BSA. 

\subsubsection{Hypersound Attenuation}
Figure \ref{fig:viscoelas} (D) shows hypersound attenuation for all solutions at various solute concentrations determined from Equation \ref{eq:fabs}.  The curves that appear in this figure were obtained in a manner analogous to those for apparent viscosity. The attenuation increases monotonically with increasing concentration for all solutions. As was the case for apparent viscosity, hypersound attenuation in the Ficoll 70 solution is approximately equal to that in the Ficoll 400 solution at low concentrations. As concentration increases, the solutions transition from the dilute to the semi-dilute and the attenuation values for the two solutions diverge. This is an indication that hypersound attenuation is much more strongly correlated with D$_{2}$O than with Ficoll in the dilute limit. In the semi-dilute limit, the dependence on Ficoll concentration becomes more important. Furthermore, similar to viscosity, at low concentrations, where attenuation is more closely related to the solvent, attenuation of BSA solutions is greater than that of either Ficoll solution. Once again, however, the concentration dependence of BSA solutions with concentration is less sharp than that of Ficoll.

\begin{table*}[t]
\caption{Empirical equations describing the concentration dependence of the elastic and viscoelastic properties of aqueous solutions of Ficoll 70, Ficoll 400, and Bovine Serum Albumin (BSA). }
\begin{ruledtabular}
\begin{tabular}{cccc}
Quantity & Base Equation & Solution & Solution-Specific Equation \\ \hline
\multirow{3}{*}{\shortstack{Hypersound Frequency\\(GHz)}} & \multirow{3}{*}{$f(x) = f_R(x) + A_1x^2$} & Ficoll 70-D$_2$O & $f(x) = (6.80/\sqrt{1-0.00653x})+0.000412x^{2}$ \\
 & & Ficoll 400-D$_2$O &   $f(x) = (6.82/\sqrt{1-0.00771x})+0.000314x^{2}$ \\ 
 & & BSA &  $f(x) = (7.45/\sqrt{1-0.00665x})+0.000253x^{2}$ \\
 \hline
 \multirow{3}{*}{\shortstack{Brillouin Linewidth\\(GHz)}} & \multirow{3}{*}{$\Gamma_B(x) = Af(x)^2 + Bx + A_2x^2$} & Ficoll 70-D$_2$O &  
 $\Gamma_{B}(x) = 0.0123x + 0.000566x^{2}+ 0.0108 f(x)^{2}$ \\
  & & Ficoll 400-D$_2$O & 
   $\Gamma_{B}(x) = 0.00899x + 0.000753x^{2} + 0.0108 f(x)^{2}$ \\ 
  & & BSA & 
   $\Gamma_{B}(x) = 0.00896x + 0.000446x^{2} + 0.0124 f(x)^2$\\
\end{tabular}
\end{ruledtabular}
\label{tab:elas_vscoel_eqs}
\end{table*}

\begin{table}[t]
\caption{Estimated overlap concentration for aqueous solutions of Ficoll 70, Ficoll 400, and Bovine Serum Albumin (BSA) obtained in present work and previous studies. PW-C: Present Work - Overlap concentration estimated from visible change in behaviour observed in plot of Frequency shift vs. Solute Concentration and/or FWHM vs. Solute Concentration. PW-Q: Present Work - Overlap concentration estimated from the divergence of previous model resulting from removing quadratic term from Eq. \ref{eq:freqshift}. } 

\begin{ruledtabular}
\begin{tabular}{cccccc}
\multirow{2}{*}{Solution} & \multicolumn{5}{c}{Overlap Concentration (wt. \%)} \\
 & PW-C & PW-Q & Ref \cite{rang2022} & Ref \cite{gtar2017}  & Ref \cite{acos2017} \\ \hline
Ficoll 70 & $\sim15$ & $\sim10$ & 10-15 & - & 22.9\\
Ficoll 400 & $\sim17$ & $\sim20$ & 10-15 & 5.33  & 13.5\\
BSA & $\sim13$ & $\sim20$ & - & - & - \\
\end{tabular}
\end{ruledtabular}
\label{tab:fit_params2}
\end{table}

\subsection{Central Mode: Relaxation}

\subsubsection{Origin}
Figure \ref{fig:fasymm} shows the central peak present in spectra of high concentration Ficoll solutions. Such peaks were not observed in BSA spectra. This peak was observed in the spectra of solutions with $x\geq20$\%, from which it was noted that its intensity increases with increasing concentration for both Ficoll 70 and Ficoll 400 solutions. The width of this central peak, however, shows no systematic change with change in concentration. This central peak was attributed to a relaxation mode due to D$_{2}$O molecules within the hydration shells of Ficoll, a phenomenon which has been observed in Brillouin and Rayleigh scattering experiments on other high concentration macromolecular solutions \cite{tao1993light,tao1989reorientational, lee1993brillouin,orecchini2009collective}. This phenomenon occurs when D$_{2}$O molecules briefly bind to the hydration shell of the Ficoll molecules before returning to the bulk solvent. This relaxation time is, therefore, also considered the residence time for solvent molecules within the hydration shell \cite{garcia1993computation}.

\begin{table}
\caption{Hydration relaxation time for aqueous Ficoll solutions (present work) and other aqueous macromolecular solutions.}
\label{tab:relaxation}
\begin{ruledtabular}
\begin{tabular}{ccc}
\multirow{2}{*}{Solution} & Macromolecule & Relaxation \\
 & Structure & Time (ps) \\ \hline
35 wt\% Ficoll 70 - Pres Work & Globular & 22 $\pm$ 2 \\
35 wt\% Ficoll 400 - Pres Work & Globular & 18 $\pm$ 2 \\
DNA \cite{tao1993light} & Chain & 40 \\
Hyaluronic Acid \cite{lee1993brillouin}& Chain & 50 \\
Polyethylene glycol 600 \cite{poch2006struc} & Chain & 60\\
\end{tabular}
\end{ruledtabular}
\end{table}

\subsubsection{Relaxation Time}
Relaxation times associated with the central peaks were determined from the linewidths using the relationship  $\tau = 1/\pi \Gamma_C$ \cite{kojima2012precursor}.  These times were found to be $22\pm 2$ ps and $18 \pm 2$ ps for the Ficoll 70 and Ficoll 400 solutions, respectively, and are representative of the duration that D$_{2}$O molecules spend within the hydration shells of Ficoll before returning to the bulk solution.  Table \ref{tab:relaxation} compares these times to hydration relaxation times for other macromolecular solutions. As shown, the relaxation time for both varieties of Ficoll solution are lower than those for aqueous solutions of DNA or hyaluronic acid \cite{tao1993light,lee1993brillouin}.  This is a reasonable result because relaxation time is directly proportional to polymer chain length and Ficoll is spherical in shape \cite{rang2022} while DNA and hyaluronic acid are long chain polymers \cite{pinnow1968rayleigh}. It should also be noted that Ranganathan et al \cite{rang2022} found that the NMR relaxation rates had reached their saturation value by concentration of 35\%, with Ficoll-400 having higher rates than Ficoll-70. This is consistent with the smaller hydration timescale observed here.

\subsection{System Properties and Dynamics}
Table \ref{tab:elas_vscoel_eqs} contains a summary of the dependence of hypersound frequency and Brillouin peak linewidth on concentration given by Equations \ref{fig:ffreqFic} and \ref{eq:FWHM}.  All elastic and viscoelastic properties explored in the present study increase monotonically with increasing solute concentration. These increases can be attributed to an increase in packing of solute molecules within the solution. For aqueous solutions of Ficoll or BSA,  there are two components which impact the volume fraction. The first is the natural addition of solute molecules to solution as concentration increases. The second component to consider is the hydration of solute molecules. This process is seen to primarily occur in higher concentration solutions, but makes a large contribution to the volume fractions of Ficoll 70 and Ficoll 400. A change in volume fraction due to hydration also occurs in BSA, but to a lesser degree than in Ficoll. As such, there is a much more dramatic increase in volume fraction of these macromolecules compared to solutions where there is no hydration of solute molecules. Because of this, Ficoll solutions used in this study range from dilute solutions to nearly maximum packing of randomly distributed spheres, and BSA solutions also experience a very high degree of packing at the highest concentrations. From the Brillouin scattering results, it is apparent that BSA experiences less packing compared to Ficoll. Table \ref{tab:elas_vscoel_eqs} shows fit equations for frequency versus solute concentration. The second order term for BSA, corresponding to polymer-polymer interactions, is $\sim$20\% lower than that of either type of Ficoll. This increase in volume occupied by globular molecules results in a reduction in solution compressibility and a corresponding increase in bulk modulus. Furthermore, the increase in solute volume fraction makes the aqueous solution more viscous and the packing of spheres increases the attenuation of sound within the solution.

To further discuss the effects of hydration, above 15\% concentration a relaxation mode was observed in Brillouin spectra for both Ficoll 70 and Ficoll 400 solutions. This relaxation was caused by hydration of Ficoll, specifically due to an exchange of D$_{2}$O between the bulk solvent and the hydration shell of Ficoll. This addition of relaxation due to hydration further validates the larger solute volume fraction due to hydration, compared to bare solute at similar mass concentrations, by demonstrating a fundamental change in the mechanics of the solution as concentration surpasses the overlap concentration. 

\section{Conclusion}
 In the present study Brillouin light scattering experiments were performed on solutions of Ficoll 70 and Ficoll 400 dissolved in D$_{2}$O with concentrations ranging from 1\% to 35 wt\%, and BSA dissolved in phosphate buffer with concentrations ranging from 0\% to 27\%. Brillouin spectra for all such solutions exhibited a single Brillouin peak which was attributed to a longitudinal bulk mode. The frequency shifts and linewidths of these Brillouin peaks were used to calculate hypersound velocity, attenuation, and bulk modulus, all of which exhibited an increase with increasing concentration. For the solutions studied in this work, the relationship between hypersound frequency and solute concentration cannot accurately be described by models which have been previously presented for non-interacting macromolecular solutions.  A model was derived to describe the change in hypersound frequency in macromolecular solutions which incorporates solute-solute interactions. 

Finally, a central peak was observed in high concentration spectra for both Ficoll 70 and 400 but was not observed in BSA spectra. This peak was attributed to relaxation associated with hydration of Ficoll by D$_{2}$O, and the occupation time of D$_{2}$O within the hydration shell.  The widths of this central peaks were used to calculate relaxation times of $\sim22$ ps and $\sim18$ ps for Ficoll 70 and Ficoll 400 solutions, respectively.  

This work provides physical insight into the interaction of macromolecules and water in crowded macromolecular environments, which are of immense importance in biological systems. Further, it provides quantitative spectroscopic signatures for bound or surface-associated water. Concentration dependence, with complementary work on temperature dependence, of aqueous biomacromolecular solutions is important because of the wide range of solution concentrations found in naturally occurring biological systems. This work also further establishes Brillouin spectroscopy as a valuable probe of the elasticity and viscoelasticity of aqueous biomacromolecular systems.  

\begin{acknowledgments}
AY and GTA acknowledge the support of the Natural Sciences and Engineering Research Council of Canada (RGPIN-2019-04970 and RGPIN-2015-04306, respectively).
\end{acknowledgments}

\bibliographystyle{apsrev4-2}
\bibliography{thesisbib}

\end{document}